\documentclass[a4paper,useAMS]{iau-JDSS}
\usepackage{color,graphicx}
\title[Stream and spopradic meteoroids associated with Near Earth Objects] 
{Stream and sporadic meteoroids associated with Near Earth Objects}

\author[T.J. Jopek and I.P. Williams]   
{Tadeusz J. Jopek$^1$,  %
  Iwan P. Williams$^2$%
 }
\affiliation{$^1$Astronomical Observatory Institute, Faculty of Physics, A.M. University, Poznan, Poland 
\\ [\affilskip]
$^2$School of Physics and Astronomy, Queen Mary University of London, E1 4NS, UK 
}
\pubyear{2013}
\volume{Volume 16}  
\pagerange{119--126}
\date{?? and in revised form ??}
\jname{Highlights of Astronomy, Volume 16}
\editors{Montmerle T. prime ed.}
\begin{document}

\maketitle
\vspace{-6pt}
\begin{abstract}
NEO’s come close to the Earth's orbit so that any dust ejected from them, might be seen as a meteor shower. Orbits evolve rapidly, so that a similarity of orbits at one given time is not sufficient to prove a relationship, orbital evolution over a long time interval also has to be similar. 
Sporadic meteoroids can not be associated with a single parent body, they can only be classified as cometary or asteroidal. However, by considering one parameter criteria, many sporadics are not classified properly therefore two parameter approach was proposed. 
\vspace{-6pt}
\keywords{meteors, meteoroids, meteoroids, methods: numerical}
\end{abstract}
\vspace{-12pt}
\section{Introduction}
%

The association between meteoroid streams and comets is well-established, though the idea that some originate from asteroids is quite old (see Williams, 2011). 
The formation mechanism of a meteor stream from a comet is well-understood. Ices sublimate generating a gas outflow that carries away small dust grains.  Inter-asteroid collisions, as was seen in 2010A2 (Linear) (Jewitt {\it et al.}, 2010) is the most obvious formation process, though there are other possibilities. In all cases, the ejection velocity is small compared to the orbital speed so that a stream is formed on an orbit that is similar to that of the parent (the physics and mathematics involved are given in Williams, 2004). 

Individual meteoroid orbits can evolve in a significantly different way from the parent. The obvious mechanism causing this is a close planetary encounter 
(Hughes Williams \& Fox, 1981) but collisions between stream meteoroids, the parent or interplanetary dust particles can also play a part. These processes feed the sporadic population, making it hard to associate such meteoroids with a given parent. 


The major tool for determining a pairing of parent and meteoroid stream has been orbit similarity.
To quantify this, Southworth \& Hawkins (1963) and Drummond (1981) proposed orbital similarity D-criteria, both being widely used. 
Using such criteria, many pairings between comets and meteor streams have been established and also between minor showers and near-Earth asteroids (NEAs). Because of the
vast increase in the number of known asteroids, the probability of a chance similarity of orbits is high. 
For this reason, Porub{\v c}an, Korno{\v s} \& Williams (2004) proposed that orbits need to remain similar for 5000 years before any association can be claimed. 
 
Three major showers, the Geminids, the Quadrantids and the Taurids all have asteroids associated with them. Whipple (1983) pointed out the orbital 
similarity between the Geminid stream and (3200) Phaethon, while Fox, Williams \& Hughes (1983) showed  that Phaethon could be the parent of the Geminid meteoroid stream assuming the ejection mechanism was Cometary. Battams \& Watson (2009) reported that Phaethon brightened by at least 2 magnitudes in 2009. The asteroids 2005 UD and 1999 YC have orbits that are similar to that of Phaethon, all suggesting that a fragmenting large comet may be the actual parent.

Jenniskens (2004) noted the similarity between the orbits of the Quadrantids 
and asteroid 2003 EH1(now 196256).
In the activity curve of the Quadrantids there is a broad background and a sharp narrow central peak suggesting that an old comet (probably 96P/Machholz), produced the background while a recent outburst produced the sharp peak. This could the fragmentation of C/1490 Y1 a few hundred years ago (Williams {\it et al} 2004), producing asteroid 2003EH1. 

The Taurid stream is a complex of several smaller streams and filaments. There are many associated asteroids and lists are 
given in Porub{\v c}an, Korno{\v s} \& Williams (2006). The Taurid complex has an active comet associated with it, 2P/Encke. Asher, Clube \& Steel (1993) suggested that the family of meteoroid streams, comet 2P/Encke and associated Apollo asteroids could all have formed by the fragmentation of a giant comet 20-30 Ky ago. 
\vspace{-15pt}
\section{Sporadic meteoroids}
The sporadic meteors have lost the identity of their parent's orbit.
In order to discriminate between the orbits of comets and asteroids, several one parameter criteria have been proposed, they are listed in Jopek \& Williams (2013). 
They have been used by several authors: Whipple, 1954, found that cometary types dominated within photographic meteors, Jones \& Sarma, 1985; Steel, 1996, found that the proportion of cometary and asteroidal were about equal within TV meteors; according to Steel, 1996 asteroidal meteoroids dominated among Kharkov radio meteors, while Voloshchuk {\it et al.}, 1997, claimed the opposite.

Applying several criteria  to 780 comets and 7830 NEAs, Jopek \& Williams (2013) found that the Q- and E-criterion were the most reliable.
They also classified $\sim 78000$ sporadic meteoroids collected from several sources. Using one parameter Q- (or E-) criterion the authors found that amongst all sporadic meteoroids  $22$-$23$\% can not be classified correctly. The orbits of such objects have asteroidal aphelia $Q<4.6$ (or semi-major axes $a <2.8$) and cometary inclinations $i>75$ deg. The ecliptic radiants of the meteors corresponding to these objects concentrate around the position of the apex of the Earth motion. Such meteoroids as was shown by Jones et al. 2001, Nesvorny et al. 2011 can be of cometary origin. 
Therefore to classify the sporadic meteoroids Jopek \& Williams proposed that two parameter criteria such as $Q$-$i$ are used. For cometary orbits we have: 
\begin{equation}
 Q=a(1+e) > 4.6 \rm{[AU]}, \hspace{0.3cm}  \rm{or} \hspace{0.3cm}
 i > 75\deg
\label{QIcriterion}
\end{equation}
where: $a$ --- semi-major axis; $e$ --- eccentricity; $i$--- inclination of the meteoroid orbit. 

%
%

\end{document}